\documentclass[aps,prl,preprint,amsmath,amssymb,showpacs]{revtex4}
\usepackage{graphicx}
\begin{document}
\title{
First Measurement of Proton-Proton Elastic Scattering at RHIC}
\date{\ \today}
\author{ S.~B\"{u}ltmann}      
\author{I.~H.~Chiang}
\author{R.~E.~Chrien}
\author{A.~Drees}
\author{R.~L.~Gill}
\author{W.~Guryn}
\author{D.~Lynn}
\author{C.~Pearson}
\author{P.~Pile}
\author{A.~Rusek}
\author{M.~Sakitt}
\author{S.~Tepikian}
\affiliation{Brookhaven National Laboratory, Upton, NY 11973, USA}
\author{J.~Chwastowski} 
\author{B.~Pawlik}
\affiliation{Institute of Nuclear Physics, Cracow, Poland}
\author{M. Haguenauer}
\affiliation{Ecole Polytechnique, 91128 Palaiseau Cedex, France}
\author{A.~A.~Bogdanov}
\author{S.~B.~Nurushev}
\author{M.~F.~Runtzo}
\author{M.~N.~Strikhanov}
\affiliation{Moscow Engineering Physics Institute, Moscow, Russia}
\author{I.~G.~Alekseev}
\author{V.~P.~Kanavets}
\author{L.~I.~Koroleva} 
\author{B.~V.~Morozov} 
\author{D.~N.~Svirida}
\affiliation{Institute for Theoretical and Experimental Physics, Moscow, Russia}
\author{M.~Rijssenbeek} 
\author{C.~Tang}
\author{S.~Yeung}
\affiliation{Stony Brook University, Stony Brook, NY 11794, USA}
\author{K.~De}
\author{N.~Guler}
\author{J.~Li}
\author{N. Ozturk}
\affiliation{University of Texas at Arlington, Arlington, TX 76019, USA}
\author{A.~Sandacz}
\affiliation{Institute for Nuclear Studies, Warsaw, Poland}

\begin{abstract}

The first result of the pp2pp experiment at RHIC on elastic scattering of
polarized protons at $\sqrt{s} = 200$ GeV is reported here. The
exponential slope parameter $b$ of the diffractive peak of the elastic
cross section in the $t$ range $0.010 \leq |t| \leq 0.019$ (GeV/c)$^2$
was measured to be \linebreak
b  =  16.3 $\pm 1.6\, ({\rm stat.}) \pm 0.9\, 
      ({\rm syst.})\, (\rm{GeV/c})^{-2} \, .$

\end{abstract}
\pacs{13.75Cs, 29.27Hj, 14.20Dh}
\maketitle


Although elastic scattering has been measured in $p \overline{p}$
collisions up to $\sqrt{s}$ = 1.8 TeV, the highest energy $pp$ data
reach only to 63 GeV. We present here the first measurement of the
slope parameter $b$ in forward proton-proton elastic scattering
obtained by the pp2pp experiment at the Relativistic Heavy Ion
Collider (RHIC) at $\sqrt{s} = 200$ GeV.

The pp2pp experiment \cite{guryn} is designed to measure polarized $pp$
elastic scattering at RHIC, which will provide proton beams with
polarizations of 0.7 and luminosities up to $2 \times 10^{32}\,\mbox{
cm}^{-2}\mbox{sec}^{-1}$. The main goal of the experiment is to study
the spin dependence of elastic scattering in the squared four-momentum
transfer range $4\times 10^{-4}$
$\leq |t| \leq 1.3$ (GeV/c)$^{2}$ and 50
$\leq\sqrt{s}\leq 500$ GeV.

By measuring elastic scattering of polarized protons in the
nonperturbative regime of QCD at RHIC, one has a unique opportunity to
probe the spin structure of the nucleon and of the exchanged
mediators of the force, the Pomeron and its odd C-parity partner, the
Odderon. The pp2pp experiment, part of the RHIC spin program, 
studies the physics of 
elastic scattering
and diffractive dissociation. It  addresses  the main 
unsolved problems in particle physics-- long range QCD and confinement.

The slope $b$ for $|t| \le 0.5$\, (GeV/c)$^{2}$ is inherently sensitive to the exchange
process, and its dependence on $\sqrt{s}$ will allow to distinguish
among various QCD based models of hadronic interactions. Some
interesting features of $b$ observed in $p \overline{p}$ are not
yet confirmed in $pp$ elastic scattering. In general, the forward peak
does not show a simple exponential behavior. The $t$
distribution becomes less steep as $|t|$ increases from 0.02
(GeV/c)$^{2}$ to 0.20 (GeV/c)$^{2}$, although at the highest Tevatron
energies this was not observed.  It is therefore of interest to see
the $b$ behavior in the RHIC energy range.

In RHIC the two protons collide at the interaction point (IP), and
since the scattering angles are small, scattered protons stay within
the beam pipe of the accelerator. They follow trajectories
determined by the accelerator magnets until they reach the detectors,
which measure the $x, y$ coordinates in the plane perpendicular to the
beam axis. The coordinates are related by the beam transport equations
to the corresponding quantities at the IP:
\begin{eqnarray}
x&=&a_{11} \cdot x_0 + L_{eff}^x \cdot \theta_x^{*} +a_{13} \cdot
y_0 + a_{14} \cdot \theta_y^{*} \nonumber \\
y&=&a_{31} \cdot
x_0+a_{32} \cdot \theta_x^{*}
 + a_{33} \cdot y_0 + L_{eff}^y \cdot \theta_y^{*} 
\label{trans}
\end{eqnarray}
where $x_0$, $y_0$ and $\,\theta_x^{\ast}$, $\theta_y^{\ast}\,$ are
the positions and scattering angles at the IP and $a_{ij}$ and $L_{eff}$
are the elements of the transport matrix. The optimum condition
for the experiment is to minimize the dependence of the measured
coordinates on the unknown collision vertex, i.~e.~to have the
$a_{ij}$'s small and the $L_{eff}$'s as large as possible.  
In that case, called ``parallel to point focusing'', rays
that are parallel to each other at the interaction point are
focused nearly to a single point at the detector.
Since in practice such a
condition is achieved for one coordinate only, in our case $y$,
Eq.~(\ref{trans}) then simplifies to 
$y \approx L_{eff}^y \cdot \theta_{y}^{\ast}$.

\begin{figure}
\includegraphics[width=90mm]{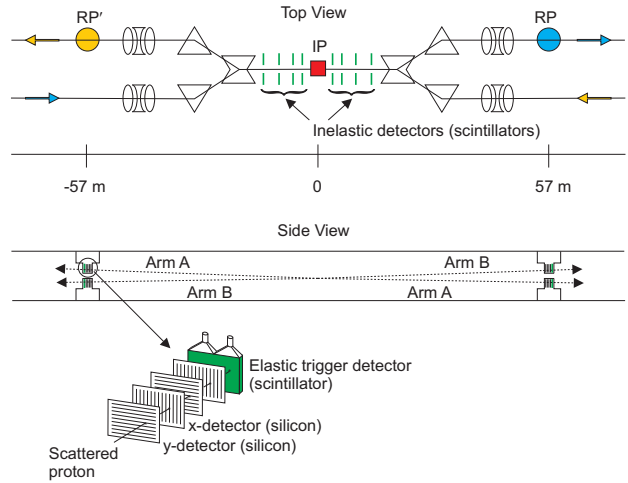}
\caption{\label{layout} Layout of the pp2pp experiment. Note the
detector pairs RP and RP$^\prime$ lie in different RHIC rings. 
Scattering is detected
in either one of two arms: Arm A is formed from the upper half
of RP$^\prime$ and the lower half of RP. Conversely, Arm B is formed from
the lower half of RP$^\prime$ and the upper half of RP.}
\end{figure}

The momentum-transfer interval for the data presented here is
$0.004 \leq |t| \leq 0.032$ (GeV/$c$)$^2$. In our 14 hour run of
January 2002, the RHIC orbit betatron function \cite{harrison} at the IP
was $\beta^* = 10$ m, resulting in $L_{eff}^y \approx 24$ m.
At larger momentum
transfers the acceptance is limited by the aperture of the RHIC
focusing quadrupoles.

The layout of the experiment is shown in Fig. 1. The
identification of elastic events is based on the collinearity 
criterion, hence it requires the simultaneous detection of the scattered
protons in the pair of Roman Pot (RP) detectors \cite{battiston}, RP and 
RP$^\prime$, on either side of the IP. 
Additionally, a set of scintillators located
outside of the beam pipe near the IP provide detection of inelastic
events.

The RP's are insertion devices allowing four
silicon strip detectors (SSD) to be positioned just above and
below the beam orbits. The SSD's inside the pots record the 
$x, y$ coordinates of the scattered protons.
The silicon detectors are made of 0.40 mm thick n-type silicon with $p^+$-type
implanted strips of 0.07 mm width and a strip pitch of 0.10 mm.
Two of the detectors have 512 strips implanted along the longer
side of the rectangle, the other two 768 strips perpendicular,
resulting in an active area of $75 \times 45$ mm$^2$.
Each strip is capacitively coupled to an input channel of a
SVXIIe \cite{lipton}, which has 128 channels with preamplification, a 32
event pipeline, and a Wilkinson-type ADC.



The amount of charge collected due to a 100 GeV/c proton
passing through the silicon detector corresponds to an energy 
deposit of about
200 keV. In 80\% of the events, this deposited energy is confined to
a single strip, and otherwise shared between neighboring strips if
the particle passed through a $30 \mu$m wide region in between the
strips. 

The elastic trigger scintillators were 8 mm thick,
$80 \times 50$ mm$^2$ in area, and were viewed by two photomultiplier tubes.
To produce a highly efficient and uniform trigger
the two signals from the tubes formed a logical OR. 
The elastic event trigger is a coincidence between signals
in the RP's scintillators, belonging either to arm A or arm B
(see Fig. 1).
The trigger efficiency was greater than 0.99. 
For each event, time and amplitude were digitized and recorded. 


The coordinate in the SSD is calculated as an energy-weighted average
of the positions of the hit strips.
Clusters of more than three hit strips were excluded. 
The detection
efficiency for every SSD strip was calculated using the redundancy
of the silicon planes for identification of elastic events. 
The average silicon detector plane efficiency for arm A was 0.97.

\begin{figure}[t]
\includegraphics[width=90mm]{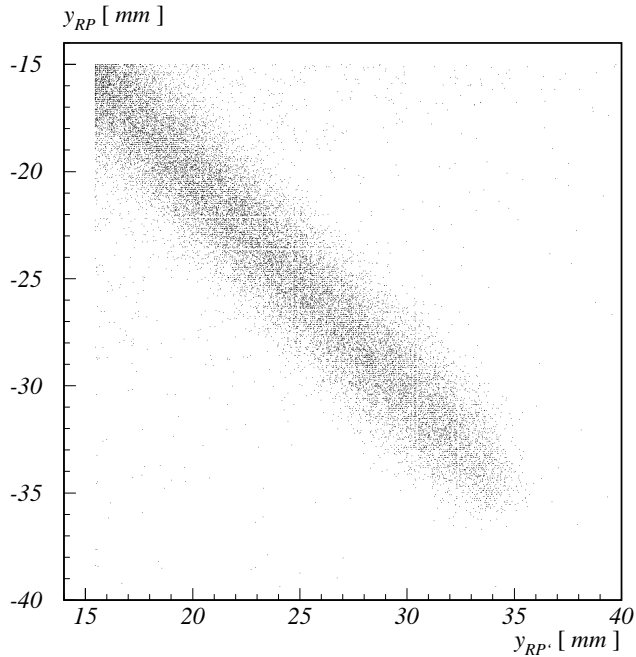}
\caption{\label{events} Correlation between the $y$ coordinates
as measured by the two detectors of arm A for elastic events
before cuts being applied.}
\end{figure}

The collinearity of elastic events implies that the 
two coordinates obtained from the silicon detectors
on either side of the interaction point are correlated. 
This correlation is shown for the $y$ coordinates 
from arm A in Fig.~\ref{events}.
The widths of the coordinate difference distributions, 
$\sigma_x$ and $\sigma_y$, were determined.
Events for which 
$ \sqrt{ \Delta x^2 + \Delta y^2 } 
   \leq 4 \, \sqrt{ \sigma_x^2 + \sigma_y^2 } $
were retained for the analysis.
The widths are dominated by the beam angular emittance of
about 12$\pi$ $\mu$m and by the uncertainty of about 60 cm (rms) in
the vertex position along the beam axis.

At least six of the possible eight planes were required to have hits
to be accepted for elastic events.
Out of 196,000 elastic triggers for arm A about 84\% were reconstructed.
Most reconstruction failures are accounted for by the larger area of
the scintillator compared to the active area of the SSD packages. The
above mentioned correlation cut of 4$\sigma$ removed another 3.8\%,
while the requirement of six hit planes contributing to the track
reconstruction cut another 0.3\%. To reduce the contamination of the
elastic event sample with tracks from background particles, not more
than two planes with more than one hit per event were accepted.  This
reduced the event sample by another 3.2\%, giving a total of 153,000
elastic events for this arm.  
A similar analysis was carried out for arm B, 
but because of the noise level being considerably
higher, it was used only for consistency checks, but not
included in the final analysis presented here.


\begin{figure}[t]
\includegraphics[width=90mm]{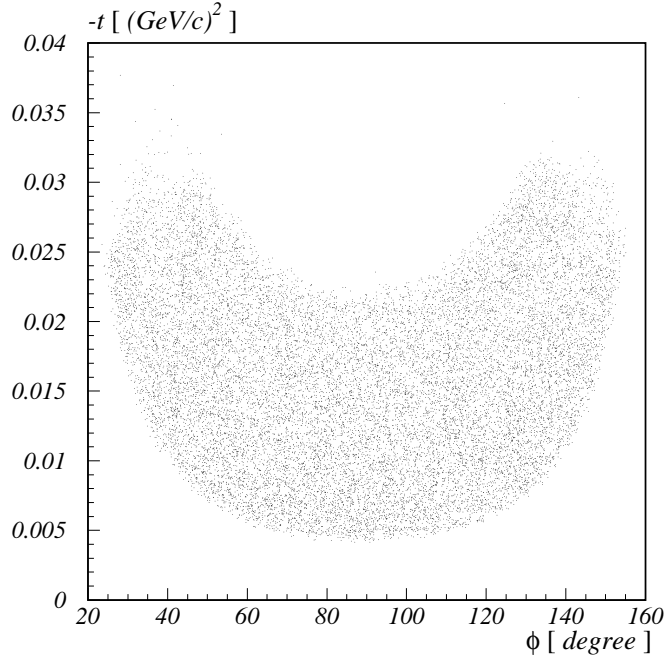}
\caption{Correlation between $t$ and
$\phi$ for reconstructed events.}
\label{phicut}
\end{figure}

\begin{figure}[t]
\centering \includegraphics[height=85mm]{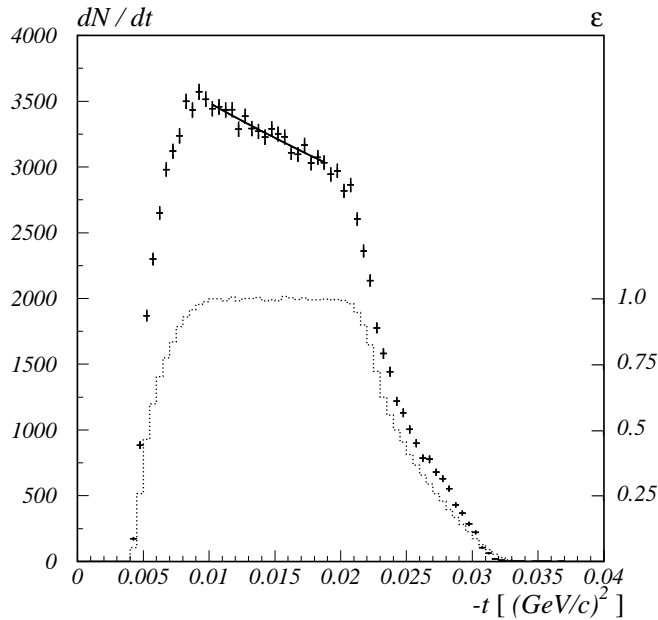}
\caption{ The distribution of d$N$/d$t$ within the $\phi$ region
selected as described in the text. The two distributions shown are the
measured data and the simulated acceptance function. The fit is
shown by the solid line.}

\label{tdist}
\end{figure}

For each event the scattering angle $\theta$ and azimuth $\phi$
were calculated for each proton and then averaged.
The scattering
angle is related to the square of the four-momentum transfer, $t$, via
$-t \approx ( \,p \cdot \theta \,)^2$.  A restriction of the $\phi$
range leads to a uniform geometric acceptance in a limited
$t$-range.  For $45^\circ < \phi < 135^\circ$ or $225^\circ < \phi
< 315^\circ$ that range is 0.010 $\leq |t| \leq $0.019 (GeV/c)$^{2}$. 
In Fig. \ref{phicut} the correlation between $t$ and
$\phi$ is shown for reconstructed events.
The determination of the slope parameter $b$ is
confined to the $t$ region for which no acceptance correction is
required. The final selection therefore yields 58,511 events.
The uncorrected d$N$/d$t$ distribution resulting from the $\phi$-cut 
is shown in Fig. \ref{tdist} together with the 
acceptance function obtained from Monte Carlo studies.

The differential cross section d$\sigma$/d$t$ for elastic scattering
in the forward angle region is determined by Coulomb and nuclear
amplitudes and the interference term between them. 
The cross section is given by (see for example Ref.~\cite{amaldi})

%
\begin{eqnarray}
\frac{{\rm d} \sigma}{{\rm d} t} & = &
4\, \pi\, (\hbar c)^2 \left( \frac{\alpha \, G_E^2}{t} \right)^2 \nonumber \\
& + & \frac{1 + \rho^2}{16 \, \pi \, (\hbar c)^2} \cdot \sigma_{tot}^2
\cdot e^{-b \,|t|} \nonumber \\
 & -  & (\rho + \Delta \Phi) \cdot \frac{\alpha \, G_E^2}{|t|} 
\cdot \sigma_{tot} \cdot  e^{-\frac{1}{2}b \,|t|} \, ,
\label{dndt}
\end{eqnarray}
with $\alpha$ the fine structure constant, 
$G_E$ the electric form factor of the proton, 
$\Delta \Phi$ the Coulomb phase\cite{kop},
$\rho$ the ratio of the real to imaginary part of the forward
scattering amplitude, 
$\sigma_{tot}$ the total cross section,
and $b$ the nuclear slope parameter. 
The dominant contribution in our $t$ region is the second
term in this expression. 

A least squares fit was performed to the distribution of Fig. \ref{tdist}
using Eq.~(\ref{dndt})
with $b$ and a normalization constant as free parameters. 
The total cross section and $\rho$ were fixed to 
$\sigma_{tot} = 51.6$ mb \cite{donnachie} and $\rho = 0.13$ \cite{augier}.
These values of $\sigma_{tot}$ and $\rho$ come from
fits to the existing $pp$ data taken at energies below 63 GeV  
and world $p \overline{p}$ data.

The resulting slope parameter is
\[
b  =  16.3 \pm 1.6\, ({\rm stat.}) \pm 0.9\, 
      ({\rm syst.})\, (\rm{GeV/c})^{-2} \, .
\]

The evaluation of the systematic errors due to the uncertainty in
beam emittance, vertex positions and spread, beam transport matrix elements,
and incoming beam angles was based on Monte Carlo simulations.
These simulations used the geometry of the experimental setup 
and efficiency of the detectors as an input.
The largest single source of the systematic error was the
uncertainty of the initial colliding beam angles.

There is also a correlation between $b$ and the values
of $\rho$ and $\sigma_{tot}$.
We find that the changes $\Delta \rho = \pm0.02$ and 
$\Delta \sigma = \pm4$ mb result in changes in
$b$ of $\Delta b = \mp 0.32$ (GeV/$c)^{-2}$
and $\Delta b = \mp 0.07$ (GeV/$c)^{-2}$, respectively.

An independent analysis of the data was performed 
using different selections of hits and elastic events.
In particular, a $t$-dependent cut on $\phi$ was
applied, which allowed an increase in the $t$ range
and the number of accepted elastic events. The $b$ slope
values obtained from both analyses agree within
statistical errors.

\begin{figure}[Ht]
\includegraphics[width=90mm]{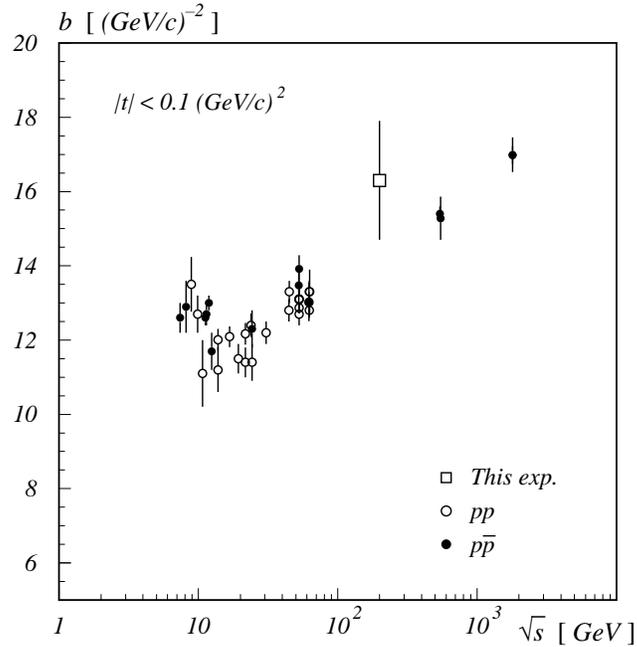}
\caption{The result for the slope parameter $b$ of this experiment
compared to the world $pp$ and $p \overline{p}$ data set. The data are
drawn from the Durham Database Group (UK).  Only statistical errors
are shown.}
\label{data}
\end{figure}

Our result for the slope parameter $b$ is shown in Fig.~\ref{data}
together with the world data on elastic $pp$ and $p \overline{p}$
scattering. This result is about one standard
deviation higher than an extrapolation of world data to the energy of
this experiment \cite{donnachie}, \cite{block}, \cite{kopel}.

In the future, a full complement of two sets of Roman Pot
detector pairs will be used, two pairs at each side of the IP, 
to allow a direct measurement of the scattering angles. 
This will reduce the systematic error due to the uncertainty
of the vertex position.
An expected increase of the RHIC luminosity
will result in a reduction of the statistical error and will
make the studies of the polarized 
observables $A_N$ and $A_{NN}$ feasible.

The research reported here has been performed in part under the US DOE
contract DE-AC02-98CH10886, and was supported by 
the US National Science Foundation and the Polish Academy of Sciences.
The authors are grateful for the help of N. Akchurin, 
D. Alburger, Y. Onel, A. Penzo, and P. Schiavon at an early stage 
of the experimental design and the support of the BNL Physics Department, 
Instrumentation Division, and C-A Department at the RHIC-AGS facility.


\begin{thebibliography}{99}
\bibitem{guryn} W.~Guryn {\em et al.}, RHIC Proposal R7 (1994) (unpublished);
V.~Kanavets, Czech.~J.~Phys., Suppl.~A, {\bf 53}, A21 (2003)
\bibitem{harrison} M.~A.~Harrison, The RHIC Project, Fifth European Particle
Accelerator Conference, Geneva (1996)
\bibitem{battiston} R.~Battiston {\em et al.}, Nucl.~Instr.~Meth.~{\bf
A238}, 35 (1985)
\bibitem{lipton} R.~Lipton, Nucl.~Instr.~Meth.~{\bf A418}, 85 (1998)
\bibitem{amaldi} U.~Amaldi {\em et al.}, Phys.~Lett.~{\bf B43}, 231 (1973)
\bibitem{kop} B.~Z.~Kopeliovich and A.~V.~Tarasov,
Phys.~Lett.~{\bf B 497}, 44 (2001).
\bibitem{donnachie}A.~Donnachie and P.~V.~Landshoff, 
Phys.~Lett.~{\bf B296}, 227 (1992)
\bibitem{augier} C.~Augier {\em et al.}, Phys.~Lett.~{\bf B315}, 503 (1993)
\bibitem{block} Martin M. Block, Nuc. Phys {\bf B} (Proc. Suppl.) {\bf 71}
378 (1999)
\bibitem{kopel} B. Z. Kopeliovich, I. K. Potashnikova, B. Povh, 
and E. Pedrazzi, Phys. Rev. {\bf D63} 054001 (2001) 
\end{thebibliography}
\end{document}